\documentclass[conference,letterpaper]{IEEEtran}
\usepackage[hyphens]{url}
\usepackage{hyperref}
\hypersetup{breaklinks=true, colorlinks,allcolors=blue}
\usepackage[backend=biber,style=ieee,citestyle=numeric-comp]{biblatex}
\IEEEoverridecommandlockouts

\usepackage{amsmath,amssymb,amsfonts}
\usepackage{algorithmic}
\usepackage{graphicx}
\usepackage{textcomp}
\usepackage{xcolor}
\usepackage{tikz}
\usepackage{array}
\usepackage{balance}
\usetikzlibrary{arrows.meta,calc}

\begin{document}

\title{Governing Bring Your Own AI:\\ A Parameterized Maturity Model}

\author{\IEEEauthorblockN{Dare Bello, John Hastings}
\IEEEauthorblockA{\textit{The Beacom College of Computer \& Cyber Sciences} \\
\textit{Dakota State University}\\
Madison, SD, USA \\
dare.bello@trojans.dsu.edu, john.hastings@dsu.edu}
}

\maketitle

\begin{abstract}
Employees are increasingly using personally owned generative AI tools such as ChatGPT, Gemini, and Claude for their daily work. This practice is known as Bring Your Own AI (BYOAI), which is a distinct form of Shadow AI in which employee-authenticated personal accounts are used outside of enterprise identity and security controls. Existing frameworks were designed for AI tools managed by organizations, and their coverage does not extend to unmanaged AI tools used with a personal account. In addressing these issues, we developed a governance model through a systematic review of the literature that produces a risk taxonomy and a framework-engagement profile. We also developed a parameterized governance model that measures how much a level of governance maturity reduces residual risk. A five-level maturity ladder is coupled to a technical control architecture through a chain in which the coverage of the control layer influences the security outcomes. Our study of a curated corpus of 30 records (24 research studies and 6 framework documents) indicated that the most prominent categories identified were data exposure and compliance, and framework engagement was inconsistent. Three mutually supporting pillars (technical, governance, and human) were established to support safeguards. Additionally, the results of the model demonstrated that prohibition-based solutions will result in residual risk levels close to those achieved through baseline solutions. Under the specified parameterization, layered control-based solutions substantially reduce modeled exfiltration risk and increase enforceable coverage.
\end{abstract}


\section{Introduction}
Personal generative AI tools have become part of daily work, and employees have adopted personal systems such as ChatGPT, Gemini, and Claude to draft content, summarize documents, help with coding, and support analysis. This trend is similar to previous trends from Bring Your Own Device (BYOD) and Shadow IT, in which employees used their own device or technology before formal governance was implemented \cite{b1,b2}. The use of employee-driven technology usually emerges due to perceived inadequacies in organizational resources, such as speed, availability, or alignment with actual workflow requirements \cite{b3,b4}.

Shadow AI, the unauthorized or unmonitored use of AI within organizations, has gained attention as a growing concern among many stakeholders with documented risks around data leakage, misaligned model behavior, and accountability gaps \cite{b2,b5,b6}. Unlike earlier forms of Shadow AI, BYOAI represents a new form of shadow AI because it involves the use of personal accounts owned by employees and authorized by employees that operate outside of organizational identity systems, network monitoring, and security controls \cite{b5,b7}. These personal accounts on unapproved AI tools create variable output that is shaped by unknown training data, thereby introducing bias, hallucinations, or factual inconsistencies into work products \cite{b8}. In addition, they can inadvertently expose sensitive organizational data to external systems \cite{b5,b9}.

The current frameworks for managing AI related risks, including the NIST AI Risk Management Framework, various responsible-AI guidelines, and AI Trust, Risk and Security Management (AI TRiSM) controls, provide valuable conceptual and technical guidelines \cite{b10,b11,b12}. However, they were developed with AI owned by an organization in mind and do not directly address the visibility and control gaps associated with BYOAI. Prior BYOAI research work generally provides conceptual perspectives and has tended to present governance ratings as asserted values rather than as outputs of a stated, reproducible process. This paper addresses both gaps by developing a governance model grounded in a systematic literature review and expressing its scoring through a parameterized model whose numbers are generated by a script rather than assigned manually.

The contribution is threefold: (i) an evidence-based BYOAI risk taxonomy and a framework-engagement profile derived from a systematic literature review; (ii) a five-level governance maturity ladder coupled with a technical-control architecture; and (iii) a parameterized governance model that measures how governance maturity flows through a chain of control coverage, security outcomes, and residual framework gap, supplemented by a retrospective mapping of four previously documented real incidents and one healthcare-sector pattern. The model is defined as directional and deterministic rather than empirical as parameters are literature-derived where sources exist and stated as explicit assumptions otherwise, and empirical validation is left to future work. Our study is guided by four research questions:
\begin{itemize}
\item \textit{RQ1 (Risk):} Which risk categories most accurately characterize BYOAI within organizations?
\item \textit{RQ2 (Frameworks):} How completely do existing governance frameworks cover BYOAI risks, and where do they fall short?
\item \textit{RQ3 (Safeguards):} Which technical and governance controls most reduce residual BYOAI risk?
\item \textit{RQ4 (Maturity):} How does a parameterized maturity model quantify an organization's progression from restrictive to optimized governance?
\end{itemize}

\section{Background and Related Work}
BYOAI builds on previous patterns of employee-driven technology adoption. BYOD has demonstrated the rapid entry of personal devices into workplace ecosystems prior to having effective governance in place, raising questions about data ownership and device management \cite{b3}. Shadow IT expanded on this by introducing the use of unmonitored cloud-based applications that were used to bypass rigid approval processes \cite{b1}. Shadow AI has been shown to demonstrate the same types of concerns associated with shadow IT in terms of risks, including unverified datasets, unreliable models, unregulated exposure to sensitive data, and inconsistent access controls \cite{b2,b5,b6}. Sector-specific research has been conducted within the healthcare, financial services, and education sectors and each has shown that the use of unregulated AI poses an increased level of risk due to the level of regulatory and confidentiality requirements \cite{b5}.

BYOAI differs because the accounts are personal rather than corporate. An organization typically will have limited insight into what their employees share with external AI systems, how those systems store or reuse the data, or whether the generated output influences decision-making, which could go against established company policies. Because AI systems transform, store, and reuse information rather than merely transmitting it, the governance implications extend beyond device management to model behavior, output reliability, and long-term retention by third-party providers. Employee-led adoption is also motivated by sociotechnical mismatch: employees turn to personal tools when approved systems are slow, unavailable, or not suitable for the task, and prohibition without a usable alternative tends to convert visible experimentation into concealed use \cite{b3,b4,b13}.

Existing frameworks provide foundational elements, but not coverage. The NIST AI RMF 1.0 outlines lifecycle-oriented risk management for AI within organizational control \cite{b10}. AI TRiSM framework provides model monitoring, runtime inspections, and adversarial defense, but assumes visibility into model operations that personal accounts do not provide \cite{b12}. Responsible-AI and ethical frameworks emphasize fairness and transparency, but focus on institutional AI development rather than personal use \cite{b11,b14}. Previous work specific to BYOAI introduces governance policies and control domains but stops short of quantifying framework alignment or control coverage \cite{b5,b7}.

The BYOAI-Gov framework of Anthuvan \textit{et al.} \cite{b1} derived from a 45-article systematic review and a 345-respondent multi-regional survey, is the most similar previous study. In contrast to this paper, it defines BYOAI as a governable ``workplace behavior'' and provides a behavior-aware framework with three components: user archetypes; task-risk zoning; and an organizational maturity layer. Our study and theirs are related but distinct. While BYOAI-Gov provides governance postures for organizations based on an author synthesis of what will be effective, we provide a reproducible quantification of the amount of residual risk that remains at each maturity level, using a deterministic, parameterized chain --- technical-control coverage score (TCCS), to security outcomes, to a Framework Gap Index (FGI), to a Composite Governance Score (CGS). Central to this is our FGI, which measures how much of a framework's nominal coverage stays enforceable under BYOAI: because AI frameworks assume organization-owned systems, only a fraction of that coverage holds when accounts are employee-owned, a dimension BYOAI-Gov does not model. We therefore treat the larger body of behavioral evidence presented by BYOAI-Gov as supplementary to, rather than supplanting, our smaller but reproducible and quantified control-coverage model.

\subsection{Technical Controls Relevant to BYOAI Governance}
Most of the concrete BYOAI risk surface appears where data and access leave the organization's visibility, so data, identity, and cloud-facing controls matter as much as policy. Seven control families anchor the technical pillar.

\textit{Data Security Posture Management (DSPM)} continuously maps where sensitive information resides and how it moves across cloud services, letting security teams spot outbound flows to AI endpoints, tie them to classification rules, and flag assets at risk of being copied into external models.

\textit{Cloud Access Security Brokers (CASB)} sit between users and cloud services to observe and control traffic, flagging or blocking unapproved GenAI destinations and enforcing rules when content labeled confidential is pushed to third-party AI tools.

\textit{Data Loss Prevention (DLP)} inspects content as it leaves endpoints, mail systems, or web gateways; when employees paste source code, client records, or internal documents into personal AI tools, DLP can alert, require justification, or block the transfer.

\textit{SaaS Security Posture Management (SSPM)} targets how SaaS applications are configured rather than individual transactions, surfacing the misconfigured sharing settings, over-permissive integrations, and unmanaged AI plug-ins through which BYOAI often appears.

\textit{Identity and Access Management (IAM)} and \textit{Cloud Infrastructure Entitlement Management (CIEM)} govern who can reach what data and under what conditions; they do not control personal AI accounts directly, but tighter scoping of permissions shrinks the pool of high-value information available to be copied into external models.

\textit{Secure Web Gateways (SWG)} enforce restrictions on specific AI domains known to store or reuse user-submitted content. Above these, runtime inspection and AI TRiSM functions---monitoring inputs and outputs and checking for abnormal model behavior---inform the upper maturity levels, where AI interactions are treated as auditable events rather than opaque tools.

Early maturity levels use only a few of these families; higher levels progressively add more and assume they operate in combination.

\section{Research Methodology}
This study combines a systematic literature review (SLR) with a design-science artifact. The review establishes an evidence-based risk taxonomy and a framework-engagement profile for BYOAI; the model then uses these findings, together with stated modeling assumptions, to demonstrate how governance maturity affects residual risk. Because BYOAI activity occurs largely outside enterprise logging, primary telemetry is scarce; a review-plus-model design is therefore appropriate, and every reported value is traceable either to the coded corpus or to an explicitly labeled assumption.

\subsection{Search Strategy}
Three sources were searched: IEEE Xplore, Scopus, and Google Scholar with a structured Boolean query for semantic discovery. The query combined three concept blocks with AND, with synonyms within each block combined using OR: (i) the phenomenon (``shadow AI,'' ``BYOAI,'' ``bring your own AI,'' ``unsanctioned AI,'' ``unauthorized AI,'' ``shadow IT''); (ii) the technology (``generative AI,'' ``large language model,'' ``LLM,'' ``GenAI,'' ``agentic AI''); and (iii) the governance angle (governance, risk management, compliance, data protection, oversight). The searches were restricted to the title, abstract, and keyword fields and to 2018--2026.

\subsection{Inclusion and Exclusion Criteria}
Inclusion criteria were defined before selection. A record was included if it addressed unapproved, personal, or shadow use of AI in an organizational context, or a governance framework or technical control directly applicable to that context; was a peer-reviewed article, conference paper, a scholarly working paper, or substantive report from a recognized body (e.g. NIST, OWASP, ISO); was published 2018--2026; and was available in English. Records were excluded if they were purely technical AI/ML papers with no governance dimension, Shadow-IT papers with no AI element (unless used solely as behavioral grounding), non-substantive sources (blogs, vendor marketing), duplicates, or papers where AI use was incidental.

\subsection{Screening and the Corpus}
\begin{figure}[htbp]
\centering
\resizebox{\columnwidth}{!}{
\begin{tikzpicture}[
  font=\footnotesize,
  box/.style={draw,rectangle,align=center,inner sep=5pt,text width=5cm,minimum height=1.1cm},
  excl/.style={draw,rectangle,align=center,inner sep=4pt,text width=3.3cm,minimum height=1cm},
  stage/.style={draw,fill=black!12,align=center,minimum width=0.6cm,minimum height=2.4cm,rotate=90,font=\footnotesize\bfseries},
  arr/.style={-{Latex[length=2.2mm]},thick}
]
\node[box] (id)   at (0, 9)  {Records identified for BYOAI relevance through database searching (IEEE Xplore, Scopus, Google Scholar) and citation chasing};
\node[box] (scr)  at (0, 6.4){Records screened by title, abstract, and full text against predefined inclusion criteria (Section III-B)};
\node[box] (elig) at (0, 3.9){Reports assessed for eligibility (full text retrieved)};
\node[box,fill=black!5] (inc) at (0, 1) {Studies included in review (N = 30): 24 research studies + 6 framework/standard documents; risk-eligible M = 24};
\node[excl] (ex1) at (6.2, 5.15){Records excluded: off-topic to BYOAI governance; non-substantive sources};
\node[excl] (ex2) at (6.2, 2.55){Reports excluded (n = 1): endpoint-detection paper, not BYOAI governance};
\draw[arr] (id) -- (scr);
\draw[arr] (scr) -- (elig);
\draw[arr] (elig) -- (inc);
\draw[arr] (0,5.15) -- (ex1.west);
\draw[arr] (0,2.55) -- (ex2.west);
\node[stage] at (-3.6, 9)   {Identification};
\node[stage] at (-3.6, 5.15){Screening};
\node[stage] at (-3.6, 1)   {Included};
\end{tikzpicture}
}
\caption{PRISMA-style flow of records through identification, screening, and inclusion.}
\label{fig:prisma}
\end{figure}
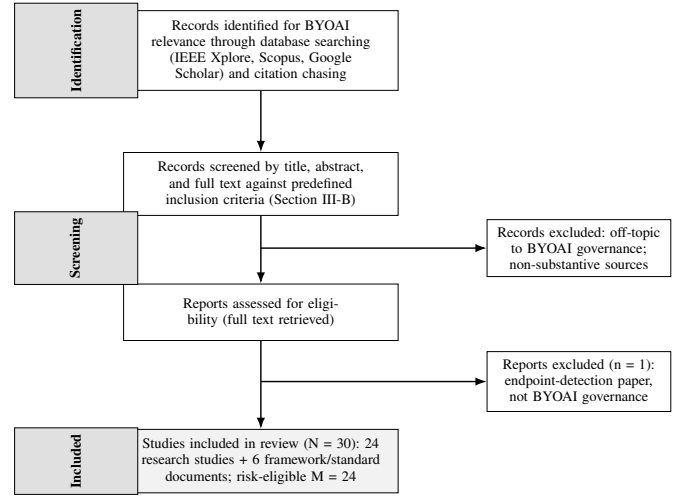

IEEE Xplore, Scopus, and Google Scholar were used to search for candidate articles, and it was supplemented with backward and forward citations from previously identified papers. The articles were then compiled into a curated body of work based on relevance to BYOAI. Since our initial search was performed in a larger context of working literature, we did not report an unfiltered database yield; instead, identification is defined as the set of BYOAI-relevant records carried forward for screening. Candidate records were initially screened using their titles, abstracts, and full texts, excluding those deemed irrelevant, off-topic, or non-substantive. The screening yielded 30 included records, comprising 24 research studies and 6 framework or standard documents evaluated as objects of analysis, as shown in the PRISMA-style flow (Fig.~\ref{fig:prisma}). The complete review corpus is represented in \cite{b1,b2,b3,b4,b5,b6,b7,b8,b9,b10,b11,b12,b13,b14,b15,b16,b17,b18,b19,b20,b21,b22,b23,b24,b25,b26,b27,b28,b29,b30}. We acknowledge as a limitation that because identification was drawn from a curated collection rather than a single reproducible database export, exhaustive-search completeness cannot be claimed.

\subsection{Coding Procedure}
The extraction process for each study identified within this systematic review was coded using an organized extraction form with respect to the six categories contained within the BYOAI Risk Taxonomy and the four framework families(NIST AI RMF, AI TRiSM, ethical/responsible AI frameworks and technical controls).

\textit{Rule 1, Risk coding.} A risk category was coded as present only when a study substantively examined that construct, not when it was mentioned in passing. A study could be included for behavioral or contextual value, but could not contribute to the risk-frequency counts.

\textit{Rule 2, Framework coding.} A framework was coded only when a study substantively evaluated, applied, compared, or operationalized it; a mere citation was insufficient. This distinction between mentioning and using a framework yields a more defensible and reproducible basis for the framework analysis.

\textit{Reporting consequence.} Because standards and framework documents are evaluated as objects of analysis rather than risk studies, they are excluded from the risk-frequency denominator. Of the $N=30$ included records, $M=24$ were risk-eligible; risk frequencies (Table~\ref{tab:risk}) are reported against $M$, while the framework engagement(Table~\ref{tab:fw}) is reported against $N$.

\section{Review Findings}
\subsection{BYOAI Risk Categorization Frequencies (RQ1)}
Table~\ref{tab:risk} reports how frequently each risk category was substantively examined in the 24 risk-eligible studies. Data exposure and privacy leakage dominate, appearing in well over half the corpus, followed by a cluster of compliance, governance-drift, and hallucination risks. Intellectual-property and bias concerns appear less frequently, often embedded within broader discussions of data protection or ethics. This distribution reflects where the reviewed literature concentrates its attention; it indexes research emphasis rather than real-world incidence.

\begin{table}[t]
\caption{BYOAI Risk Category Frequency ($M=24$ Risk-Eligible Studies)}
\begin{center}
\begin{tabular}{|l|c|c|}
\hline
\textbf{Risk Category} & \textbf{\# Studies} & \textbf{\% of M} \\
\hline
Data Exposure / Privacy & 14 & 58 \\
\hline
Compliance / Legal & 11 & 46 \\
\hline
Governance Drift / Visibility & 10 & 42 \\
\hline
Hallucination / Output Risk & 9 & 38 \\
\hline
IP / Confidentiality & 6 & 25 \\
\hline
Bias / Fairness & 6 & 25 \\
\hline
\end{tabular}

{\footnotesize \raggedright \% of M = proportion of the 24 risk-eligible studies coding each category.\par}

\label{tab:risk}
\end{center}
\end{table}

\subsection{Framework Engagement (RQ2)}
Table~\ref{tab:fw} reports how many included studies substantively engaged each framework family, applying Rule 2. Technical controls and ethical/responsible-AI frameworks are engaged most often, while AI TRiSM is engaged by comparatively few studies and NIST AI RMF by a moderate number. This uneven engagement indicates that the literature treats no single framework as a complete answer to BYOAI, and that the frameworks most oriented toward model internals are least represented in a setting where the model is not organization-owned. The magnitude of this gap---not just its unevenness---is quantified as the Framework Gap Index in Section~\ref{sec:model}, where the enforceable share of nominal framework coverage under BYOAI falls to roughly one quarter.

\begin{table}[t]
\caption{Framework Engagement Across Included Studies ($N=30$)}
\begin{center}
\begin{tabular}{|l|c|c|}
\hline
\textbf{Framework} & \textbf{\# Studies} & \textbf{\% of N} \\
\hline
Technical Controls & 9 & 30 \\
\hline
Ethical / Responsible AI & 8 & 27 \\
\hline
NIST AI RMF & 6 & 20 \\
\hline
AI TRiSM & 3 & 10 \\
\hline
\end{tabular}

\vspace{1pt}
{\footnotesize \raggedright \% of N = proportion of all 30 included records engaging each framework.\par}

\label{tab:fw}
\end{center}
\end{table}

Table~\ref{tab:fw} reports framework \textit{engagement} (whether a study substantively evaluates or applies a framework), a direct coding output. The finer framework-by-risk coverage matrix used as an input to the model (Section~\ref{sec:model}) is an analytical assessment defined by the authors informed by the reviewed literature, and is labeled as such rather than presented as a coding result.

\section{Governance Artifact and Parameterized Model}\label{sec:model}
\subsection{Three-Pillar Control Model (RQ3)}
\label{sec:three_pillar_model}
The reviewed evidence confirms that BYOAI cannot be governed solely by any single form of control. Therefore, we organize safeguards into three complementary pillars. The technical pillar (DSPM, CASB, DLP, SSPM, IAM, CIEM, SWG) governs where data and access leave organizational visibility. The governance pillar provides policy and legal or contractual controls (e.g., vendor no-training clauses and business associate agreement gating) as well as providing the necessary mechanisms to assign accountability and to provide human-in-the-loop verification. The human pillar addresses the sociotechnical drivers of adoption (role-specific and workflow embedded training, sanctioned enablement that creates compliant use as the most convenient option, near-miss reporting, and creating a culture in which compliant use is modeled rather than penalized). The three pillars are interdependent since each addresses types of risk that the other two pillars do not: technical controls prevent data egress but not fabricated output; governance assigns accountability but it does not change behaviors; and human controls reduce the incentive to reach for personal tools in the first place.

The ladder advances all three pillars in parallel, summarized in Table~\ref{tab:pillars}. Early levels rely on policy or informal guidance alone, leaving the drivers of BYOAI unaddressed, while Levels 3--4 combine technical controls with governance mechanisms such as human-in-the-loop verification and human-pillar enablement. Level 5 integrates all three pillars under continuous oversight and a reporting-and-learning culture that reduces the incentive for BYOAI rather than simply blocking it.

\begin{table}[t]
\caption{Maturity Levels Mapped to the Three Control Pillars}
\begin{center}
\footnotesize
\renewcommand{\arraystretch}{1.2}
\setlength{\tabcolsep}{3pt}
\begin{tabular}{|>{\raggedright\arraybackslash}p{1.35cm}|>{\raggedright\arraybackslash}p{1.95cm}|>{\raggedright\arraybackslash}p{1.85cm}|>{\raggedright\arraybackslash}p{1.7cm}|}
\hline
\textbf{Level} & \textbf{Technical} & \textbf{Governance} & \textbf{Human} \\
\hline
L1 Prohibition & --- & policy / ban only & --- \\
\hline
L2 Awareness & DSPM (limited scanning) & informal guidance & --- \\
\hline
L3 Guardrails & DSPM, DLP, SWG & approved workflows & role-specific training; sanctioned pathway \\
\hline
L4 Integrated & + CASB, SSPM, IAM (runtime/TRiSM) & named accountability; HITL verification & --- \\
\hline
L5 Optimized & + CIEM; continuous monitoring, auto-enforce. & full integration & reporting-and-learning culture \\
\hline
\end{tabular}
\label{tab:pillars}
\end{center}
\end{table}

\subsection{Model Design and Parameter Provenance}
The model is deterministic and parameterized, not empirical: it shows directionality of the effect rather than measured magnitudes, without parameter distributions or repeated runs (Section~\ref{sec:limits}). We deliberately limit our focus to only the technical pillar where there exists a defendable coverage metric for its control families. The Technical Control Coverage Score (TCCS) is used to measure this. The governance and human pillars are handled qualitatively throughout this paper because their effects (training efficacy, accountability, cultural change) cannot be parameterized like the technical layer, and assigning them mitigation weights would manufacture unsupported precision. Their role is captured in the maturity ladder and the case mapping rather than in the numeric output. The model represents a causal chain where the governance level determines what control families are active; the active stack determines technical-control coverage (TCCS); coverage determines security outcomes (exfiltration risk, detection latency, attack-surface breadth); and those outcomes, with framework coverage, determine the residual FGI and composite CGS. Each control family reduces residual risk in each axis area that it primarily addresses, based on a Zero-Trust-style model that assumes no single control provides end-to-end protection. Model parameters fall into two provenance classes: values informed by published research or industry reports, and explicitly stated modeling assumptions. These include per-family mitigation weights (DLP, CASB, IAM sector studies), detection-latency anchor (breach-dwell time reporting), framework coverage assessment, and coded risk frequencies (Section IV).

\subsection{Metric Definitions}
The Technical Control Coverage Score is the share of the seven canonical control families -- DSPM, CASB, DLP, SSPM, IAM, CIEM, and SWG active at a maturity level:
\begin{equation}
\mathrm{TCCS} = \frac{\text{active control families}}{7}. 
\label{eq:tccs}
\end{equation}
Rather than nominal coverage, the Framework Gap Index measures the gap in \textit{enforceable} coverage under BYOAI. Because the frameworks target AI owned by organizations, only a fraction $\alpha$ of their nominal best-of-framework coverage is enforceable when the account is owned by employees. The baseline gap (Level~1) is
\begin{equation}
\mathrm{FGI}_0 = 1 - \frac{\alpha \cdot \text{covered}_\text{nominal}}{\text{total}}, \quad \alpha = 0.25. \label{eq:fgi0}
\end{equation}
As maturity rises, technical controls restore enforceability, closing a compensable fraction of the baseline gap in proportion to TCCS:
\begin{equation}
\mathrm{FGI}(\ell) = \mathrm{FGI}_0 \cdot \bigl(1 - 0.90 \cdot \mathrm{TCCS}(\ell)\bigr). \label{eq:fgi}
\end{equation}
This avoids a pitfall: a nominal measure would report high baseline coverage and contradict the paper's premise. The enforceability form instead yields a large baseline gap ($\mathrm{FGI}_0 \approx 0.77$, only ${\approx}23\%$ enforceable) that narrows only as controls deploy, consistent with the finding that frameworks alone do not govern AI owned by employees. The Composite Governance Score combines risk mitigation $R$ (mean fractional reduction across the three outcome axes), governance alignment $C = 1 - \mathrm{FGI}$, and technical deployment $T = \mathrm{TCCS}$:
\begin{equation}
\mathrm{CGS} = 1 + 4\,(w_1 R + w_2 C + w_3 T), \label{eq:cgs}
\end{equation}
with $(w_1,w_2,w_3) = (0.40, 0.30, 0.30)$. Because $C$ reflects enforceable alignment, an ungoverned Level~1 environment scores near the floor ($\mathrm{CGS} \approx 1.3$), which is not misleadingly high.

\subsection{Model Results (RQ4)}
\begin{figure}[htbp]
\centerline{\includegraphics[width=\columnwidth]{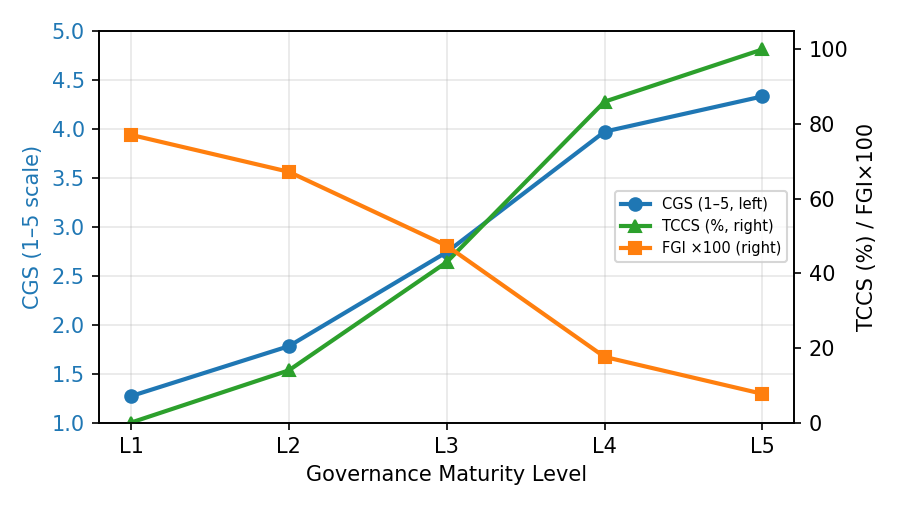}}
\caption{Governance maturity progression. CGS on its native 1--5 scale (left axis); TCCS (\%) and FGI$\times$100 share the right axis. TCCS increases by construction; CGS and FGI move accordingly across levels.}
\label{fig:maturity}
\end{figure}

\begin{figure}[htbp]
\centerline{\includegraphics[width=\columnwidth]{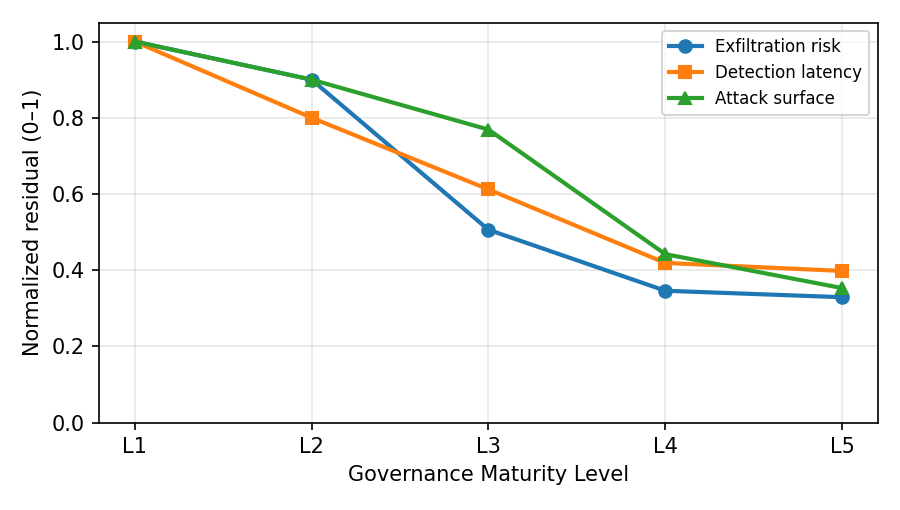}}
\caption{Security outcomes decline as maturity increases: exfiltration risk, detection latency, and attack surface (normalized).}
\label{fig:outcomes}
\end{figure}

Figures~\ref{fig:maturity} and~\ref{fig:outcomes} trace the model's behavior at the five maturity levels. By construction, stacking control families raises TCCS from 0\% to 100\% across the ladder; this rise is a design property of the maturity mapping (Table~\ref{tab:pillars}). The model's substantive output is how that coverage propagates through the chain: as TCCS rises, CGS climbs from 1.3 to 4.3 on a 1--5 scale, while residual FGI falls from 0.77 to 0.08, so that only about a quarter of the nominal framework coverage is enforceable under ungoverned BYOAI, and technical controls progressively restore it. The steepest transitions occur entering Level 3, when data-centric controls (DSPM, DLP) first come online, and Level 4, as identity and cloud-access controls integrate. Exact per-level values follow deterministically from the parameters and equations specified above.

Because the direction of these trends is structurally guaranteed rather than tuned, we swept the applicability factor $\alpha$ across $[0.15, 0.40]$ and the CGS weights across the valid simplex, including R-, C-, and T-heavy extremes. Across all combinations, FGI decreases and CGS increases monotonically at every level (Table~\ref{tab:sensitivity}); the parameters shift the magnitude of the curves but never their direction. This invariance is structural: $\alpha$ enters only through the Level-1 baseline $\mathrm{FGI}_0$ and cannot reverse the TCCS-driven decline, while any non-negative weights summing to one preserve the monotonic rise of a composite of three increasing quantities. The envelope is widest at Level~1 (FGI $0.63$--$0.86$; CGS $1.11$--$1.88$) and narrows by Level~5 (FGI $0.06$--$0.09$; CGS $4.07$--$4.66$), where deployed controls dominate the parameter choice.

\begin{table}[t]
\caption{Sensitivity Envelope of FGI and CGS across $\alpha \in [0.15, 0.40]$ and CGS Weight Vectors on the Valid Simplex (Baseline, plus R-, C-, and T-heavy Extremes). Direction is Invariant; only Magnitude Varies.}
\begin{center}
\footnotesize
\renewcommand{\arraystretch}{1.2}
\setlength{\tabcolsep}{5pt}
\begin{tabular}{|l|c|c|c|c|c|}
\hline
 & \textbf{L1} & \textbf{L2} & \textbf{L3} & \textbf{L4} & \textbf{L5} \\
\hline
FGI (min) & 0.63 & 0.55 & 0.39 & 0.14 & 0.06 \\
\hline
FGI (max) & 0.86 & 0.75 & 0.53 & 0.20 & 0.09 \\
\hline
CGS (min) & 1.11 & 1.63 & 2.61 & 3.77 & 4.07 \\
\hline
CGS (max) & 1.88 & 2.29 & 3.11 & 4.23 & 4.66 \\
\hline
\end{tabular}
\label{tab:sensitivity}
\end{center}
\end{table}

\section{Retrospective Case Mapping}
This section maps four documented incidents and one healthcare-sector pattern of non-sanctioned or personal AI use onto the risk taxonomy and maturity ladder. The analysis is \textit{retrospective and conditional}: the framework was not executed against these incidents, and no claim is made that it \textit{would have prevented} them. The cases are organized by layering thesis, technical-layer interception, governance-layer interception, and cases requiring both because the recurring lesson is that no single layer is sufficient.

\subsection{Technical-Layer Interception: Samsung (2023)}

After Samsung allowed ChatGPT use in 2023, three engineers submitted proprietary semiconductor source code,
defect-detection algorithms, and a confidential meeting transcript within roughly twenty days; the company responded with a blanket ban \cite{b35}. Although this case used an authorized rather than a private account application, it illustrates the data exfiltration process that is fundamental to BYOAI. The above case involves two risk areas: Data Exposure/Privacy and IP/Confidentiality. The organization was at maturity Levels 1 -- 2, as evidenced by their outright ban — also a Level 1 posture. Data-focused controls introduced across maturity Levels 3--4 (DSPM, DLP, SWG, then CASB) are the types of tool that can flag or block similar attempts to send such files or data. Unfortunately, the technical solution is incomplete, as engineers can find an alternative to sanctioned options simply because they can complete their task more effectively via an unauthorized tool, and banning the tool does nothing to eliminate the reason engineers chose to use it. The human pillar, a sanctioned
capable pathway plus role-specific training, is what would reduce recurrence, and its absence is why prohibition alone is the weakest posture.

\subsection{Governance-Layer Interception: Legal Sector}
Two legal-sector incidents illustrate risks that data-centric controls cannot detect. In \textit{Mata v. Avianca} (2023), an attorney filed a brief with fabricated judicial decisions produced by ChatGPT; the court imposed sanctions \cite{b31}. This engages Hallucination/Output Reliability, which DLP and CASB cannot catch because nothing sensitive leaves the organization; only a governance-layer human-verification workflow (Levels 4--5) helps, corroborating the Section IV finding that technical controls do not cover hallucination. In \textit{Heppner} (2026), a court ruled that the case-strategy materials a defendant entered into a consumer AI assistant were not protected by attorney-client privilege or work-product doctrine \cite{b32}, creating confidentiality and discoverability exposure. Here, the act of input is itself the harm; the safeguards are governance-layer usage rules and training. Together, the pair shows that the legal sector needs governance controls at both the input and output ends of AI-assisted work.

\subsection{Both Layers Required: Banking and Healthcare}
A named banking incident and a healthcare pattern require governance and technical controls together. In a community bank (2026), an employee used a personal account and device to upload customer names, Social Security numbers, and dates of birth to an unapproved AI application; the institution treated the event as material and filed an SEC Form 8-K \cite{b33}, then blocked unapproved AI domains and tightened data access. This BYOAI event exercises Compliance/Legal and Data Exposure, and its remediation reveals the required layers: domain blocking (CASB/SWG) and access tightening (IAM) alongside a governance pathway giving employees a compliant tool. Both layers must be repeated on a scale in healthcare, where personnel paste protected health information into consumer tools that generally do not sign a BAA under the Health Insurance Portability and Accountability Act (HIPAA) \cite{b34}; a BAA-gated approval process (governance) must be enforced by classification and DLP (technical), as neither suffices alone.

\begin{table}[t]
\caption{Cross-Case Mapping to Risk and the Three Control Pillars}
\begin{center}
\begin{tabular}{|l|l|l|}
\hline
\textbf{Case (sector, year)} & \textbf{Primary risk} & \textbf{Pillar(s) needed} \\
\hline
Samsung (tech, 2023) & Data Exp.; IP & Technical + Human \\
\hline
Mata (legal, 2023) & Hallucination & Governance \\
\hline
Heppner (legal, 2026) & IP/Confid. & Governance + Human \\
\hline
Comm. Bank (bank, 2026) & Compliance & Technical + Governance \\
\hline
Healthcare (pattern) & Compliance & Technical + Governance \\
\hline
\end{tabular}
\label{tab:cases}
\end{center}
\end{table}

\textit{Synthesis of three pillars.} Reading through the pillars, the cases show that the layer positioned to intercept a risk is rarely the layer that addresses its cause. Technical controls would have flagged the Samsung and banking data egress, but the behavior that produced them, employees reaching for capable personal tools under time pressure, is a human-pillar problem: survey evidence on shadow adoption reports that a majority use unsanctioned tools despite knowing the policy, and that most workplace AI users first adopt the tool outside work \cite{b1,b13}. Governance controls are the only layer that reaches the Mata and Heppner failures, which leave no technical trace. And no pillar substitutes for another: prohibition without a sanctioned pathway and training merely converts visible use into concealed use. The maturity ladder therefore advances all three pillars together, and the central finding, that effective BYOAI governance is layered, extends from two layers to three once the behavioral driver of adoption is taken seriously. All counterfactual statements are conditional descriptions of where controls would be positioned, not claims that the framework would have prevented the actual events; details are drawn from public reporting and, where available, primary artifacts (the court ruling in Mata, the SEC Form 8-K in the banking case).

\section{Discussion}
Reading through RQ1--RQ4, the review and model tell a consistent story. Data-centric risks dominate (RQ1), and four of the five mapped cases turn on regulated or proprietary data leaving organizational control. The engagement of the framework is uneven (RQ2): technical controls and ethical frameworks are used the most, AI TRiSM the least, and no framework is treated as complete. The maturity model (RQ3--RQ4) shows why layering matters: because TCCS rises by construction as controls stack, the model's substantive output is that CGS rises and FGI falls in step---but meaningfully only once controls combine, echoing Zero-Trust reasoning that no single control, and once the human pillar is included no single pillar, suffices.

Two cautions are followed. First, prohibition (Level 1) leaves residual risk near baseline because it does not address the motivations that drive personal AI use; the case evidence reinforces this, since several organizations reached bans that addressed the tool rather than the exposure pathway or the need for the underlying task. Second, partial governance provides partial protection only, and a false sense of security itself is a risk. Because FGI falls most when technical controls supplement frameworks, security teams can use CGS, FGI, and TCCS trends to justify phased, identity, and data-centric investments while preserving productivity. For practitioners, the three pillar view implies that prohibition alone is the weakest posture, that technical visibility (DSPM linking identity, classification, and cloud-access telemetry) is foundational but cannot reach output-reliability or input-confidentiality failures, and that the maturity ladder should advance all three pillars together as a portfolio rather than a menu.

\section{Limitations and Future Work}\label{sec:limits}
This model is directional and parameterized rather than empirical. It uses fixed parameters selected by the authors without distributions or repeated runs, so it is a parameterized scoring model, not a stochastic simulation; we do, however, report a sensitivity analysis over $\alpha$ and the CGS weights (Table~\ref{tab:sensitivity}) confirming the reported trends are direction-invariant. Risk frequencies derive from 24 risk-eligible studies in a curated 30-record corpus assembled from a broader collection rather than an exhaustively logged database export, so complete search coverage is not claimed. Coding was performed by a single researcher, precluding inter-rater reliability; the reported percentages therefore index relative emphasis across the corpus rather than precise population estimates, and the small denominator ($M = 24$) means individual proportions should be read as approximate. The explicit coding rules in Section~III (the present/absent decision rule and the substantive-engagement rule) constrain each judgment and make the scheme reproducible in principle; independent double-coding with a formal agreement statistic (e.g., Cohen's $\kappa$) is a priority for future work. The model quantifies only the technical pillar, while the governance and human pillars, along with the framework-by-risk coverage matrix that feeds the FGI, are treated qualitatively or as analytical input. Mitigation weights, the applicability factor $\alpha$, and the detection-latency and attack-surface axes are modeled rather than observed, and the case mapping is retrospective and conditional rather than a test of prevention. Reported values are therefore meaningful in direction and relative magnitude rather than as absolute measurements.
 Finally, the corpus is weighted toward sources published through 2023--2026, whereas organizational AI governance is evolving rapidly; some control gaps identified here may already be addressed at more mature organizations, while others remain open, so the risk landscape should be read as a snapshot of the reviewed period rather than a current-state census.

These limitations motivate the next steps. The most direct is empirical instantiation in a controlled lab, a local LLM, a simulated corporate document store, scripted employee interactions representing the risk taxonomy, and a detection layer, run across the three environments to replace assumed parameters with observed exfiltration rates, detection latencies, and attack-surface measures. Extending the sensitivity analysis to the per-family mitigation weights and to distributional rather than point parameter estimates would characterize uncertainty around the current single-point estimates. Further work includes survey and interview validation of the taxonomy, longitudinal case studies tracking organizations as they climb the ladder, adversarial and regulatory-shift scenarios, and privacy-preserving detection of personal AI use.

\section{Conclusion}
BYOAI introduces visibility gaps, compliance exposure, and inconsistent model behavior that current frameworks address only partially. This paper contributes a governance model grounded in a systematic review and expressed through a reproducible parameterized model, complemented by a retrospective mapping of four documented incidents and a healthcare-sector pattern. The evidence shows that effective BYOAI governance is cumulative and layered across three complementary pillars, technical, governance, and human. Prohibition alone leaves residual risk near baseline; integrated data- and identity-centric controls reduce exfiltration risk, shorten detection latency, and restore enforceable coverage; but only the addition of accountability structures and human-pillar enablement reaches the output-reliability, confidentiality, and adoption-driver risks that technical controls cannot.

\section*{AI Tools Used}
Overleaf's AI writing assistant was used during the preparation of this manuscript to assist in grammar and spelling.

\balance
\printbibliography

\end{document}